\RequirePackage{fix-cm}
\documentclass[twocolumn,epjc3]{svjour3}  
\smartqed  
\RequirePackage{graphicx}
\usepackage{mathtools, cuted}
\usepackage{xcolor}
\RequirePackage[colorlinks,citecolor=blue,urlcolor=blue,linkcolor=blue]{hyperref}
\journalname{Eur. Phys. J. C}
\begin{document}

\title{Stellar structure of quark stars in a modified Starobinsky gravity}


\author{Arun Mathew\thanksref{e1} \and Muhammed Shafeeque\thanksref{e2} \and Malay K. Nandy\thanksref{e3} }

\thankstext{e1}{e-mail: a.mathew@iitg.ac.in}
\thankstext{e2}{e-mail: m.shafeeque@iitg.ac.in}
\thankstext{e3}{e-mail: mknandy@iitg.ac.in}


\institute{Department of Physics, Indian Institute of Technology Guwahati, Guwahati 781039, India}

\date{Received: 10 August 2019 / Accepted: 09 June 2020}

\maketitle

\begin{abstract}

We propose a form of gravity-matter interaction given by $\omega RT$ in the framework of $f(R,T)$ gravity and examine the effect of such interaction in spherically symmetric compact stars. Treating the gravity-matter coupling as a perturbative term on the background of Starobinsky gravity, we develop a perturbation theory for equilibrium configurations. For illustration, we take the case of quark stars and explore their various stellar properties. We find that the gravity-matter coupling causes an increase in the stable maximal mass which is relevant for recent observations on binary pulsars. 

\end{abstract}

\section{Introduction} 

Modern day scenarios such as inflation \cite{Starobinsky1983,Vilenkin1985}, late-time cosmic acceleration \cite{Riess1998,Perlmutter1999,Bernardis2000}, flat rotation curves \cite{Oort1932,Zwicky1933,Smith1936,Babcock1939} etc. are incompatible with the standard prescription of general relativity (GR). Although the predictions of GR in the weak-field regime are precise,  it falls short in the higher curvature regime in the sense that it predicts singularities such as the big bang and the black hole singularities. It has been shown that quantum corrections generate higher order self-coupling curvature in addition to the original scalar curvature \cite{Davies1977,Bunch1977}. This motivates one to consider non-linear curvature theories to see if they provide a better descriptions of gravitation phenomena. 

A nonlinear curvature theory of gravity was proposed by Starobinsky \cite{Starobinsky1980} in order to address the issue of the big-bang singularity. He considered the Einstein field equations $G_{\mu\nu}=\kappa \langle T_{\mu\nu} \rangle $ where the right hand side gives quantum mechanical contributions due to coupling between quantum matter fields (having different spins) with classical gravitational field, with the assumption of  isotropy and homogeneity and absence of radiation field. In one-loop approximation, and upon regularization, $\langle T_{\mu\nu} \rangle $ was found to be  a function of the Riemann geometric quantities. Based on these findings, Starobinsky exhibited the existence of a one-parameter family of non-singular solutions of the de-Sitter type which could be analytically continued into the region $t<0$. The de-Sitter phase naturally explains the inflation scenario without having to include any inflaton field. 

However, another approach involves a generalization of the Einstein-Hilbert action where an arbitrary function $f(R)$ represents the Lagrangian density \cite{Buchdahl1970}. In the Starobinsky model, namely,  $f(R)=R+\alpha R^2$, and its other generalisations, inflation has been explained to obtain increasingly better fits the observational data \cite{Carroll2004,Artymowski2014}. Moreover, various forms of $f(R)$ gravity have been able to explain the late-time cosmic acceleration \cite{Kehagias2014,Artymowski2014}. In addition, a simple power-law form of $f(R)$ gravity is able to explain \cite{Capozziello2004,Capozziello2007} rotation curves in the spiral galaxies. The power-law form has been explored  \cite{Bernal2011,Mendoza2015,Barrientos2018} to find a basis for the modified Newtonian dynamics (MOND) which is the most successful scenario in explaining rotation curves in many different types of galaxies \cite{Milgrom1983,Bekenstein2006,McGaugh1998,McGaugh2011}. 

Inclusion of the effect of classical matter with $f(R)$ gravity came in two different forms, namely, $f(R,L_m)$ and $f(R,T)$, where $L_m$ is the matter Lagrangian and $T$ is the trace of energy-momentum tensor. While $f(R,L_m)$ gravity has been studied extensively in various contexts \cite{Harko2010,Harko2012}, $f(R,T)$ gravity entered the literature somewhat recently \cite{Harko2011}. It was noted that the $T$ dependence may arise due to exotic imperfect fluid or quantum effects. Thus it is natural to expect that $f(R,T)$ gravity may be a suitable candidate for compact objects such as neutron stars and quark stars where quantum effects are expected to play a significant role. 

Models of extended gravity have been employed to study the stellar structure of different compact objects. Starobinsky gravity with $f(R)=R+\alpha R^2$ has been applied to neutron stars treating $\alpha R^2$ as perturbation with well-known models for the equation of state \cite{Arapo_lu2011}. It was found for some cases that the maximal mass could approach $\sim2$ M$_\odot$ only for negative values of $\alpha$. Moreover  the logarithmic model, $f(R)=R+\alpha R^2 +\alpha \gamma R^2 \ln (R/\mu^2)$  \cite{Alavirad2013}, was also studied perturbatively for neutron and quark stars that exhibited similar trends for different $\gamma$ values. The Starobinsky model was further explored non-perturbatively for neutron stars \cite{Capozziello2016}. They observed that, for positive values of $\alpha$, GR yeilded higher maximum mass values than the Starobinsky case. They also studied the model $f(R) = R + \alpha R^2 (1+\gamma R)$ which exhibited low sensitivity on the $\gamma$ value. On the other hand,  for their model $R^{1-\epsilon}$,  GR gave the lowest maximum mass and the mass value increased to very high values approaching $2.5$ to $3$ M$_\odot$.

Yazadjiev et al. \cite{Yazadjiev2014} solved for the stable configurations of neutron stars in the Starobinsky model $f(R)=R+\alpha R^2$ for increasing values of the parameter $\alpha$. By constructing an equivalent scalar-tensor theory, they obtained the stellar structure non-perturbatively and compared their results with perturbative estimates.  While the perturbative result was unphysical because it gave a decreasing mass with respect to the radial distance in a region interior to the star \cite{Orellana2013}, no such unphysical behaviour was observed in the non-perturbative framework. Staykov et al. \cite{Staykov2014} included a slow rotation in neutron and strange stars in a non-perturbative framework of Starobinky gravity. While the slow rotation does not affect the mass and radius with respect to the static Starobinky case, they found a measurable increase in the moment of inertia with respect to GR.

The Starobinsky model $R+\alpha R^2$ was further studied for neutron and quark stars non-perturbatively \cite{Astashenok2017}.  For positive and non-zero values of $\alpha$, they observed that the scalar curvature does not decrease to zero at the surface (unlike the perturbative results) and it exponentially falls off outside the star. The stellar mass contribution until the surface plus the gravitational mass contribution outside the star constitute the total mass which is actually observed by a distant observer. The gravitational redshift for the distant observer will be determined by the total (stellar + gravitational) mass. The gravitational mass contribution from the outside of the star is remarkably in distinction with the perturbative approaches where the exterior solution is assumed to be Schwarzschild. For negative values of $\alpha$, they \cite{Astashenok2017} found that the Ricci scalar executes a damped oscillation beyond the surface of the star and the gravitational mass contribution increases indefinitely. In an earlier paper, the same authors \cite{Astashenok2015} compared the prediction of the Starobinsky model and the corresponding scalar-tensor theory.  Their non-perturbative analysis indicated that the star is surrounded by a dilaton sphere whose contribution to the mass is negligible. 

Models of $f(R,T)$ gravity and its generalisations were studied for equilibrium configurations of compact stars. Carvalho et al. \cite{Carvalho2017} considered the model $f (R, T ) = R -2\lambda T$ to find the equilibrium configurations for white dwarfs. The maximum mass limit obtained was slightly above the Chandrasekhar limit. In comparison to GR and $f(R)$ predictions, the white dwarfs were found to have larger radii as the $\lambda$ value was increased from zero. Deb et al. \cite{Deb2018} considered the same model to obtain the equilibrium stellar structure of quark stars. They demonstrated that the $\mathcal{M}$-$\mathcal{R}$ curves are different for positive, negative and zero $\lambda$ values.

It is important to note that, in the Starobinsky model $R+\alpha R^2$,  a maximum value of 2 M$_\odot$ or beyond is reached only when the $\alpha$ value is chosen to be negative \cite{Astashenok2017}. However, this leads to an issue, namely, the Ricci scalar executes a damped oscillation and the gravitational mass contribution increases indefinitely in the exterior region. On the other hand, the Ricci scalar smoothly decreases to zero at infinity for positive $\alpha$ values, for which the star can support a maximum mass lower than 2 M$_\odot$.  Thus a physical theory based on Starobinsky model requires a positive $\alpha$ value whence the Ricci scalar behaves properly everywhere. However, in order to reach 2 M$_\odot$ or beyond, the Starobinsky model requires modification. We therefore consider the model $f(R,T) = R+\alpha R^2 +\omega R T$ with $\alpha>0$. This modification implies inclusion of gravity-matter interaction in the description via the term $\omega R T$. It would be sufficient to show that the maximum mass attainable is greater than the Starobynsky prediction even if we take a simple form $R(1+\alpha R +\omega T)$,  and treat $\omega T$ perturbatively on the background of non-perturbative Starobinsky solution. 

In this paper, we obtain the field equations for spherically symmetric distribution of matter for $f(R,T) =R+\alpha R^2 +\omega R T$ to $\mathcal{O}(\omega)$. With this, we solve for equilibrium configurations of quark stars with the equation of state given by the bag model, namely, $p=k(\varepsilon-4B)$, where $B=60$ MeV/fm$^3$ is the bag constant, and we take the physical value $k=0.28$ which is valid for strange quark mass $m_s=250$ MeV/c$^2$. For the pure Starobinsky case,  a maximum mass of $1.832$ M$_\odot$ is obtained for $\alpha=10 r_g^2$, whereas GR gives a maximum mass of $1.764$ M$_\odot$ \cite{Astashenok2015}.  On the other hand, the present model yields a maximum mass $\sim2$ M$_\odot$, which is consistent with different observations of  binary millisecond pulsars, namely, J0348+0432, J1614-2230, and J0740+6620, with pulsar masses $2.01\pm0.04$ M$_\odot$ \cite{Antoniadis2013}, $1.93\pm{0.017}$ M$_\odot$ \cite{Demorest2010,Fonseca2016}, and $2.14^{+0.20}_{-0.18}$ M$_\odot$ \cite{Cromartie2019}, respectively. 

The paper is organised as follows. In Section \ref{Preliminary_details}, we lay out the preliminary details for the field equations and energy conservation in $f(R,T)$ gravity. In Section \ref{Present_model}, we present the details of calculation for the proposed model with gravity-matter interaction. There, we also develop a perturbative treatment as the gravity-matter interaction is expected to be small. We thus obtain the stellar equation for equilibrium configurations in spherically symmetric stars. We apply these equations to quark stars in Section \ref{QS_in_SG} and obtain the stellar properties. Section \ref{Discussion} contains a discussion on the obtained results and the main conclusions are given in Section \ref{Conclusion}.

\section{Preliminary details}\label{Preliminary_details} 

In this section we briefly present the preliminary details of $f(R,T)$ gravity needed for our later developments. The action of the most general $f(R,T)$ gravity is given by \cite{Harko2011}
\begin{equation}\label{Action}\small
S = \frac{c^3}{16\pi G} \int d^4 x \sqrt{-g} f(R,T) +  \int d^4 x \sqrt{-g} L_m
\end{equation}
where $L_m$ is the Lagrangian density of matter. The stress-energy tensor $T_{\mu\nu}$ is obtained from the matter Lagrangian $L_m$ as 
\begin{equation}\label{EMT}\small
T_{\mu\nu} = g_{\mu\nu} L_m - 2 \frac{\partial L_m}{\partial g^{\mu\nu}}.
\end{equation}

Field equations following from Eq.~(\ref{Action}) are 
\begin{eqnarray}\label{FE_1}
f_R R_{\mu\nu} -\frac{1}{2}fg_{\mu\nu} &&+ (g_{\mu\nu} \nabla^\alpha\nabla_\alpha   -\nabla_\mu\nabla_\nu)f_R \nonumber \\
&&= \kappa T_{\mu\nu}  - f_T T_{\mu\nu} - f_T \Theta_{\mu\nu}
\end{eqnarray}
where  $\Theta_{\mu\nu} = g^{\alpha\beta} \delta T_{\alpha\beta}/\delta g^{\mu\nu}$, $f_R = \partial f/\partial R$, $f_T = \partial f/\partial T$ and $\kappa=8\pi G/c^4$.

Assuming the matter to be a perfect fluid, the stress-energy tensor $T_{\mu\nu}$ can be obtained from going over to the proper frame and then switching back to the gravitational frame \cite{Landau_FDbook}, yielding  
\begin{equation}\label{EMT}
T_{\mu\nu} = (\varepsilon+p)u_{\mu} u_{\nu} + pg_{\mu\nu}, 
\end{equation}
where the energy density $\varepsilon$ and pressure $p$ are the proper values and  $u^{\mu}$ is the macroscopic four-velocity. 

It can be shown from Eq.(\ref{EMT}) that the above form of stress-energy tensor can be obtained from the choice $L_m = p$ \cite{Harkobook}. Consequently one obtains
\begin{equation}\label{Theta}
\Theta_{\mu\nu} = -2T_{\mu\nu} + pg_{\mu\nu}
\end{equation}
Thus the field equation (\ref{FE_1}) become
\begin{eqnarray}\label{}
f_R R_{\mu\nu}  -\frac{1}{2}fg_{\mu\nu} +(g_{\mu\nu} \nabla^\alpha\nabla_\alpha&& -\nabla_\mu\nabla_\nu)f_R = \kappa  T_{\mu\nu}  \nonumber \\
&& + f_T  T_{\mu\nu}  -f_T pg_{\mu\nu}
\end{eqnarray}
which can be re-written as 
\begin{eqnarray}\label{main_FE}
f_R G_{\mu\nu} -\frac{1}{2}(f-f_R R)&& g_{\mu\nu} + (g_{\mu\nu} \nabla^\alpha\nabla_\alpha -\nabla_\mu\nabla_\nu) f_R  \nonumber\\
&&= \kappa T_{\mu\nu} + f_T T_{\mu\nu} -f_T pg_{\mu\nu}
\end{eqnarray}
where $G_{\mu\nu} = R_{\mu\nu} - \frac{1}{2}g_{\mu\nu}R$ is the Einstein tensor. 

It is shown that the covariant divergence of the field equations give the identity \cite{Koivisto2006}
\begin{eqnarray}
\nabla^\mu \biggl[ f_R R_{\mu\nu} -\frac{1}{2}f g_{\mu\nu} + (g_{\mu\nu} \nabla^\alpha\nabla_\alpha -\nabla_{\mu}\nabla_{\nu})f_R \biggr] = 0
\end{eqnarray}
which in turn gives 
\begin{eqnarray}
\nabla^\mu T_{\mu\nu} = \frac{f_T}{\kappa-f_T} \biggl[ (T_{\mu\nu}+\Theta_{\mu\nu})\nabla^\mu \ln f_T + \nabla^\mu \Theta_{\mu\nu}\biggr]
\end{eqnarray}
Substituting for $\Theta_{\mu\nu}$ from Eq.~(\ref{Theta}), we obtain
\begin{equation}\small\label{}
 \nabla^\mu T_{\mu\nu}  = \frac{f_T}{\kappa  + f_T} \biggl[ (pg_{\mu\nu} -T_{\mu\nu} )\nabla^\mu \ln f_T + g_{\mu\nu} \nabla^\mu p \biggr]
\end{equation}
for a perfect fluid.

\section{Present model}\label{Present_model}

In this section we define the present model of gravity-matter interaction in $f(R,T)$ gravity. We derive the corresponding field equations, the modified TOV equations and also discuss the far-field solution. 

\subsection{Gravity-matter interaction}\label{GMI}

We consider  gravity-matter interaction in a modified gravity represented by 
\begin{equation}
f(R,T) = R +\alpha R^2 + \omega RT,
\end{equation}
where the last term represents the gravity-matter interaction. This form reduces the field equation (\ref{main_FE}) to
\begin{eqnarray}\label{FE}
\phi G_{\mu\nu}  +&& \frac{1}{2}\alpha R^2 g_{\mu\nu} + 2\alpha(g_{\mu\nu} \nabla^\alpha\nabla_\alpha -\nabla_\mu\nabla_\nu) R  \nonumber\\
&& = \kappa T_{\mu\nu} + \omega \bigg[ R (T_{\mu\nu} - pg_{\mu\nu}) - G_{\mu\nu}T \nonumber\\
&&- (g_{\mu\nu} \nabla^\alpha\nabla_\alpha -\nabla_\mu\nabla_\nu) T  \bigg]
\end{eqnarray}
where $\phi = 1 + 2\alpha R $. 

The corresponding trace equation is 
\begin{eqnarray}\label{Final_TraceEq}
6\alpha \nabla^\mu \nabla_\mu  R  + [2\alpha R&&-\phi] R  = \kappa T \nonumber \\
+&& \omega \left[  2(T-2p)R- 3 \nabla^\mu \nabla_\mu  T  \right]
\end{eqnarray}

Since the gravity-matter interaction is expected to be small, we shall take a perturbative approach about the exact solutions of $R+\alpha R^2$ by assuming $|\omega T|\ll1$. To the first order in $\omega$, we get
\begin{eqnarray}\label{FFE}
 G_{\mu\nu}  +&& \frac{\alpha R^2}{2\phi} g_{\mu\nu} + \frac{2\alpha}{\phi}(g_{\mu\nu} \nabla^\alpha\nabla_\alpha -\nabla_\mu\nabla_\nu) R  \nonumber\\
&& = \kappa \frac{T_{\mu\nu}}{\phi} + \frac{\omega}{\phi_0} \bigg[ R (T_{0\mu\nu} - pg_{0\mu\nu}) - G_{0\mu\nu}T \nonumber\\
&&- (g_{0\mu\nu} \nabla^\alpha\nabla_\alpha -\nabla_\mu\nabla_\nu) T_0  \bigg]
\end{eqnarray}
where the subscript  ``$0$'' indicates unperturbed quantities when $\omega=0$, so that $\phi_0=1 + 2\alpha R_0$. The corresponding trace equation (\ref{Final_TraceEq}) is obtained as
\begin{eqnarray}\label{Trace_Eq}
6\alpha \nabla^\mu \nabla_\mu  R  + [2\alpha R&&-\phi ] R  = \kappa T \nonumber \\
+&& \omega \left[  2(T_0-2p_0)R_0- 3 \nabla^\mu \nabla_\mu  T_0  \right]
\end{eqnarray}
up to $\mathcal{O}(\omega)$. 

Since we are interested in the spherically symmetric and static case, we assume the metric 
\begin{equation}\label{Metric}\small
ds^2 = -e^{\nu(r)}c^2 dt^2 +e^{\lambda(r)}dr^2 + r^2d\Omega^2 
\end{equation}
along with $\phi=\phi(r)$ and $T=T(r)$, and $d\Omega^2=d\theta^2 + \sin^2\theta \ d\varphi^2$.  Thus the above trace equation reduces to
\begin{eqnarray}\small\label{pre_Trace_Eq}
6\alpha&&\left[\frac{d^2}{dr^2}  + \left(\frac{\nu'}{2} - \frac{\lambda'}{2} +\frac{2}{r} \right) \frac{d}{dr}  \right]R - R  e^{\lambda}  = \kappa T  e^{\lambda}  \nonumber\\
&&+ \omega  \bigg[  2(T_0-2p_0)R_0 e^{\lambda_0} -  3 \left(\frac{\nu_0'}{2} - \frac{\lambda_0'}{2} +\frac{2}{r} \right) T_0'  \nonumber\\
&&- 3 T_0''   \bigg]
\end{eqnarray}
up to $\mathcal{O}(\omega)$. 

The $tt$-component of the field equation (\ref{FFE}) yields 
\begin{eqnarray}\label{tt_Eq}
 \lambda' =&&  \frac{1-e^\lambda}{r}  + \frac{1}{6}\frac{r e^{\lambda}}{\phi}\left( 2+ 3\alpha R \right) R -  \gamma   \nu'  \nonumber\\
 &&+ \frac{\kappa}{3}\frac{r e^{\lambda}}{\phi} \left(3 \varepsilon + T  \right)  + \omega \bigg[ \frac{re^{\lambda_0}}{3\phi_0} (3\varepsilon_0 - p_0 + 2T_0) R_0 \nonumber\\ 
 &&- \frac{1}{\phi_0}\left( r T_0' \frac{\nu_0'}{2} + T_0 \lambda_0' \right) + \frac{1-e^{\lambda_0}}{r \phi_0} T_0 \bigg]
\end{eqnarray}
up to $\mathcal{O}(\omega)$.
The $rr$-component yields the following equation
\begin{eqnarray}\label{rr_Eq}
\nu'  =&&   \frac{1}{1+\gamma}\left[ \kappa  \frac{r e^\lambda}{\phi} p + \frac{e^\lambda-1}{r}   - \frac{\alpha}{2} \frac{r e^\lambda}{\phi}  R^2 -  \frac{4}{r} \gamma \right] \nonumber \\
&&- \frac{\omega}{1+\gamma_0} \bigg[    \frac{1}{\phi} \left( T_0 + \frac{r T_0'}{2} \right)\nu_0'   +  \frac{1}{\phi}\left( \frac{1-e^{\lambda_0}}{r} \right)T_0 \nonumber\\
&& + \frac{2}{\phi} T_0'  \bigg]
\end{eqnarray}
up to $\mathcal{O}(\omega)$, where $\gamma = \frac{r}{2} (\ln \phi)'$.

\subsection{Extended TOV equation}\label{ETOV}

Covariant divergence of the field equation~(\ref{main_FE}) yields
\begin{eqnarray}\small
(\kappa+f_T)&&\nabla^\mu T_{\mu\nu} \nonumber \\
&&=f_T  \biggl[ (pg_{\mu\nu}-T_{\mu\nu}) \nabla^{\mu} \ln f_T + g_{\mu\nu}\nabla^{\mu} p \biggr]
\end{eqnarray}
Substituting $f_T=\omega R$, we obtain
\begin{equation}\label{Con_Eq}\small
 \nabla_\mu T^{\mu\nu}  = \frac{\omega R}{\kappa  + \omega R} \biggl[ (pg^{\mu\nu} -T^{\mu\nu} )\nabla_\mu \ln R + g^{\mu\nu} \nabla_\mu p \biggr]
\end{equation}

For the spherically symmetric static metric (\ref{Metric}), we obtain from Eq.~(\ref{EMT}) 
\begin{equation}\small
\nabla_\mu T^{\mu\nu} = e^{-\lambda} p' + (\varepsilon +p)\frac{\nu'}{2}e^{-\lambda}.
\end{equation}
Since Ricci scalar $R$ and pressure $p$ are functions of $r$ alone, the conservation equation (\ref{Con_Eq}) becomes 
\begin{equation}\label{}
p'    = -  (\varepsilon +p) \left( \frac{\kappa+\omega R}{\kappa}\right)\frac{\nu'}{2}
\end{equation}

To the first order in $\omega$, we obtain
\begin{equation}\label{Modified_TOV}
p'    = -  (\varepsilon +p) \frac{\nu'}{2}  -  \frac{\omega}{\kappa} (\varepsilon_0 +p_0) R_0  \frac{\nu_0'}{2}.
\end{equation}

For the case of vanishing $\omega$, we recover the original TOV equation. For  $\omega\neq0$, we designate Eq.(\ref{Modified_TOV}) as the extended TOV (ETOV) equation, where the pressure gradient depends on the value of $\omega$ as well as the Ricci scalar $R$. It is thus evident from Eq.~(\ref{Modified_TOV}) that the pressure gradient inside a spherically symmetric star will change as compared to the standard GR case. However, similarly to GR, the cumulative mass $m(r)$ is related to the metric potential $\lambda(r)$ as  
\begin{equation}\label{M}
m(r) =   \frac{c^2  r}{2G}  \bigg[ 1 - e^{-\lambda(r)} \bigg].
\end{equation}

\subsection{Far field solution}\label{Far_field_sol}

In the region exterior to the star, the trace equation (\ref{Trace_Eq}) takes the form  
\begin{equation}\label{Vacuum_Eq2}
6\alpha \nabla^\mu\nabla_\mu R - R  = 0.
\end{equation}
which is identical to that obtained for $f(R)=R+\alpha R^2$ gravity in vacuum. This suggests that the exterior solution has an  identical form in both Starobinsky gravity and for the given particular form of $f(R,T)$. 

For spherically symmetric static metric (\ref{Metric}), this equation takes the form
\begin{equation}\label{Vacuum_Eq}
e^{-\lambda}\left\{ R''+ \left(\frac{\nu'}{2} -\frac{\lambda'}{2} + \frac{2}{r}\right) R'\right\} - \frac{R}{6\alpha}  = 0
\end{equation}

We note that the Starobinsky correction $\alpha R^2$ is a very weak contribution as one approaches infinity. This is also immediately obvious from the above equation because the last term dominates in the limit $\alpha\rightarrow0$ giving us back $R=0$. Thus the choice of the Starobinsky form $R+\alpha R^2$ has to coincide with the solution of Einstein gravity at infinity. In order to see how the Einstein limit is approached at infinity, we must do an approximate analysis of Eq.~(\ref{Vacuum_Eq}). Since $\nu$ and $\lambda$ and their first derivatives are expected to approach zero on approaching infinity (also confirmed by exact numerical calculations), we can approximate Eq.~(\ref{Vacuum_Eq}) to the form
\begin{equation}\label{}
R''+ \frac{2}{r}R' - \frac{R}{6\alpha}  = 0
\end{equation}
Solution of this equation is given by
\begin{equation}\label{Vaccum_sol1}
R(r) = c_1 \frac{e^{-r/\sqrt{6\alpha}}}{r} + c_2\frac{\sqrt{6\alpha}}{2} \frac{e^{r/\sqrt{6\alpha}}}{r}
\end{equation}
Since $R\rightarrow0$ as $r\rightarrow\infty$, we have to set the integration constant $c_2=0$, giving
\begin{equation}\label{Vaccum_sol}
R(r) = c_1 \frac{e^{-\frac{r}{\sqrt{6\alpha}}}}{r}, 
\end{equation}
which approaches zero faster than $r^{-1}$ as $r\rightarrow\infty$ for positive value of $\alpha$. However, for negative values of $\alpha$, the far field solution given by (\ref{Vaccum_sol1}) is oscillatory in nature implying that negative $\alpha$ values are unphysical.

\section{Quark stars with gravity-matter coupling}\label{QS_in_SG}

In this section, we examine in detail the stellar structure of quark stars in the modified gravity model $f(R,T)= R+\alpha R^2+\omega RT$ that incorporates gravity-matter interaction. In massive compact stars (such as quark stars and neutron stars), we expect the gravitational field to be strong enough so that the gravity-matter coupling has a appreciable contribution. With the above choice of $f(R,T)$ gravity, Sections \ref{GMI} and \ref{ETOV} give a perturbative solution on the background of unperturbed Starobynsky gravity given by $f(R)=R+\alpha R^2$. The field equations (\ref{tt_Eq}), (\ref{rr_Eq}), (\ref{Modified_TOV}), together with the trace equation~(\ref{pre_Trace_Eq}), 
are reduced to a set of five first order differential equations, given by 
\begin{equation}\label{MSG_Eq0}
R' = \Psi, 
\end{equation}
\begin{eqnarray}\label{MSG_Eq1}
\Psi'     = &&-\frac{2}{r} \Psi    +\frac{\kappa}{6\alpha} T  e^{\lambda} + \frac{e^{\lambda}}{6\alpha} R  + \frac{\lambda'-\nu'}{2} \Psi  \nonumber \\
&&+ \frac{\omega }{6\alpha} \bigg[  2(T_0-2p_0)R_0 e^{\lambda_0} -  3 \left(\frac{\nu_0'}{2} - \frac{\lambda_0'}{2} +\frac{2}{r} \right) T_0' \nonumber\\
&&- 3 T_0''   \bigg],
\end{eqnarray}
\begin{eqnarray}\label{MSG_Eq2}\small
 \lambda' = && \frac{1-e^\lambda}{r}  + \frac{1}{6}\frac{r e^{\lambda}}{\phi}\left( 2+ 3\alpha R \right) R -  \gamma   \nu'  + \frac{\kappa}{3}\frac{r e^{\lambda}}{\phi} \left(3 \varepsilon + T  \right) \nonumber\\
 &&+ \ \omega \bigg[ \frac{re^{\lambda_0}}{3\phi_0} (3\varepsilon_0 - p_0 + 2T_0) R_0  \nonumber\\
 && - \frac{1}{\phi_0}\left( r T_0' \frac{\nu_0'}{2} + T_0 \lambda_0' \right) + \frac{1-e^{\lambda_0}}{r \phi_0} T_0 \bigg],
\end{eqnarray}
 \begin{eqnarray}\label{MSG_Eq3}\small
\nu'  =&&   \frac{1}{1+\gamma}\left[ \kappa  \frac{r e^\lambda}{\phi} p + \frac{e^\lambda-1}{r}   - \frac{\alpha}{2} \frac{r e^\lambda}{\phi}  R^2 -  \frac{4}{r} \gamma \right] \nonumber \\
&& - \frac{\omega}{1+\gamma_0} \bigg[    \frac{1}{\phi_0} \left( T_0 + \frac{r T_0'}{2} \right)\nu_0'   +  \frac{1}{\phi_0}\left( \frac{1-e^{\lambda_0}}{r} \right)T_0 \nonumber\\
&&  + \frac{2}{\phi_0} T_0'  \bigg],
\end{eqnarray}
\begin{eqnarray}\label{MSG_Eq4}
p'    = -  (\varepsilon +p) \frac{\nu'}{2}  -  \frac{\omega}{\kappa} (\varepsilon_0 +p_0) R_0  \frac{\nu_0'}{2},
\end{eqnarray}
where we have defined a new field variable $\Psi=R'$.

To complete the solution of the above equations, we take the equation of state of the quark star as that of quark-gluon plasma given by the bag model \cite{Jaffe1979,Simonov1981}, namely, $p=k(\varepsilon-4B)$, where $B=60$ MeV fm$^{-3}$ (or $B^{1/4}\approx147$ MeV) is the bag constant  and the value of the constant $k$ is associated with the choice of the QCD coupling constant ($\alpha_c$) and the mass ($m_s$) of strange quark; $k=0.33$ if $m_s=0$ and $k=0.28$ for the realistic value $m_s=250$ MeV/c$^2$. The values $B^{1/4}\approx147$ MeV and $m_s=250$ MeV correspond to the QCD coupling constant $\alpha_c=0$, as seen from Figure 1 in Ref. \cite{Farhi1984}.

Consistency of the perturbation theory requires $|\omega T|\ll1$ at all densities. This condition is satisfied throughout the star if one require that $|\omega T| \ll1$ is true at the center.   By defining 
\begin{equation}
\omega= \frac{\beta}{4B}, 
\end{equation}
this condition becomes $(1-3k)\beta\frac{\varepsilon_c}{4B} + 3\beta k \ll1$. For the choice of $\beta\sim10^{-2}$ and $\frac{\varepsilon_c}{4B} \sim 10$,  $|\omega T_c| \approx 4.48\times10^{-2}$, thus ensuring the validity of the perturbative approach.

We solve field equations~(\ref{MSG_Eq0})--(\ref{MSG_Eq4}) numerically upon making them dimensionless by defining  $\eta=r/r_g$, $\chi = R r_g^2$, $\chi_0 = R_0 r_g^2$, $\xi=\Psi r_g^3$, $\tilde{p}=p/4B$, $\tilde{p}_0=p_0/4B$, $\tilde{\varepsilon}=\varepsilon/4B$, $\tilde{\varepsilon}_0=\varepsilon_0/4B$, $\tilde{T}=T/4B$ and $\tilde{T}_0=T_0/4B$, where $r_g = G M_{\odot}/c^2 = 1.4766\times 10^5$ cm, is taken as the the scaling parameter. 

The numerical integrations of the above set of differential equations are carried out by requiring that the metric is asymptotically flat at infinity for the initial conditions $\lambda(0) = 0$ and $\nu(0) = \nu_c$.  Since the metric potential $\nu(r)$ enters the field equations only through its derivatives, the central value $\nu_c$ remains arbitrary and fixed by specifying a value that satisfies $\nu\rightarrow0$ for $r\rightarrow\infty$. The same initial conditions are imposed on the unperturbed metric, that is, $\lambda_0(0)=0$ and $\nu_0(0) = \nu_{0c}$. The central value of pressure $p(0)=p_c$ is assigned by the equation of state for a given central density $\rho_c$. Correspondingly, the unperturbed pressure takes the value $p_0(0)=p_c$. The surface of the star is identified at a radial distance $r_s$ (stellar radius) for which the pressure $p$ vanishes. Moreover, since the value of the scalar curvature is maximum at the center, and it gradually decreases towards the surface, we have the boundary condition $\Psi(0)=0$.

\begin{figure}[h!]
\centering
\includegraphics[width=0.48\textwidth]{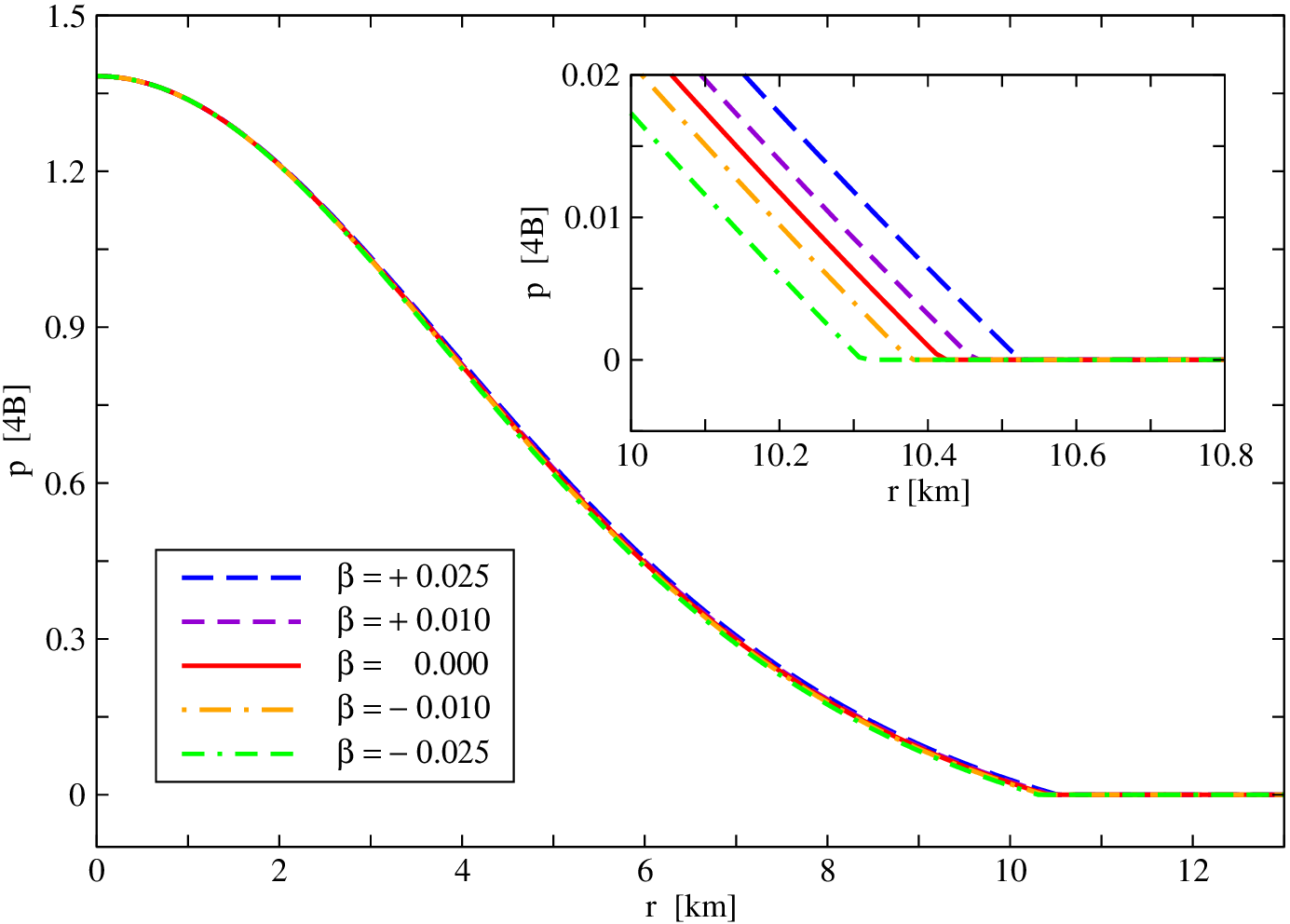}
\caption{Radial profile of pressure $p(r)$ for different values of $\beta$ ($\beta=4B\omega$) with central density $\rho_c = 2.5414\times 10^{15}$ g cm$^{-3}$. The inset shows the pressure profile near the stellar radius $r_s$  for different values of $\beta$ .}  
\label{Figure_1}
\end{figure}

A unique choice for the central value of the scalar curvature in both the unperturbed scenario (Starobinsky gravity) and the actual case requires the knowledge of the exterior solution. This requires one to continue the integration outside the star with boundary conditions $\lambda(r_s) = \lambda_s$, $\nu(r_s) = \nu_s$, $R(r_s)=R_s$ and $\Psi(r_s)=\Psi_s$ at the surface, obtained from the interior solution for an initial guess  for $R_c$. To set the above initial conditions, we first carry out a numerical integration for the unperturbed case, with similar boundary conditions  $\lambda_0(r_s) = \lambda_{0s}$, $\nu_0(r_s) = \nu_{0s}$, $R_0(r_s)=R_{0s}$ and $\Psi_0(r_s)=\Psi_{0s}$,  imposed for an initial guess $R_{0c}$. 

For convenience, the initial guess in both the cases are taken to be the GR value $^ER_c=\kappa (\rho_c c^2 -3p_c)$ since this value is not too far from the required values. The integration is carried out several times for different initial guesses until the required conditions $R\rightarrow0$ and $|\Psi|\rightarrow0$ as $r \rightarrow \infty$ and $R_0\rightarrow0$ and $|\Psi_0|\rightarrow0$ as $r \rightarrow \infty$ are satisfied. This procedure gives the correct central values $R_c$ and $R_{0c}$, which are found to be lower than the GR value. For the sake of accuracy, we fine-tune $\chi_c$ ($=R_c r_g^2$) and $\chi_{0c}$ ($=R_{0c} r_g^2$) up to twelve decimal figures by taking the exterior solution as far as $\eta_{\rm max}$ ($=r_{\rm max}/r_g$), where $\eta_{\rm max}$ satisfies $\chi(\eta_{\rm max})\sim10^{-12}$ and $\chi_0(\eta_{\rm max})\sim10^{-12}$. 

In the original framework of general relativity, the Ricci scalar vanishes immediately outside the surface, the trace equation being $R = -\kappa T$. In our present case, the vacuum solution is given by  Eq.~(\ref{Vacuum_Eq2}) and the Ricci scalar does not vanish but decays exponentially outside the star, as also implied by the far-field solution, Eq.~(\ref{Vaccum_sol}). This gives rise to two distinct masses \cite{Astashenok2015}, namely, the  stellar mass $M_s = m(r_s)$, the mass within the stellar radius $r_s$, and the mass $M$ as seen by a sufficiently distant observer,  estimated as 
\begin{equation}\label{M}
M =   \frac{c^2}{2G}  r_{\rm max}  \left\{ 1 - e^{-\lambda (r_{\rm max}) } \right\}
\end{equation}
Since the numerical calculations are sufficiently accurate with a sufficiently large value of $r_{\rm max}$, this estimate for $M$ is expected to be close to the one for $r\rightarrow\infty$. 

In the following subsections, we analyze the exact numerical solutions of the field equations given by  Eqs.~(\ref{MSG_Eq0})--(\ref{MSG_Eq4}) with the boundary conditions discussed above for quark stars with the equation of state given by the bag model. We also compare these results with the Starobinsky case, $\omega=0$.  We take $\alpha = 10r_g^2 = 2.1804\times10^{11}$ cm$^2$ (that is, $\sqrt{\alpha}=  3.16 r_g = 4.6694\times10^{5}$ cm),  which is smaller than the estimated upper bound $\sqrt{\alpha}<7\times 10^{7}$ cm as predicted by binary pulsar data \cite{Naf2010}. 

\subsection{Interior and exterior solutions}\label{I_E_sol}

In this sub section we elaborate upon the interior and exterior solutions by looking at the radial profiles for pressure $p(r)$, mass $m(r)$ and Ricci scalar $R(r)$.
\begin{figure}[h!]
\centering
\includegraphics[width=0.48\textwidth]{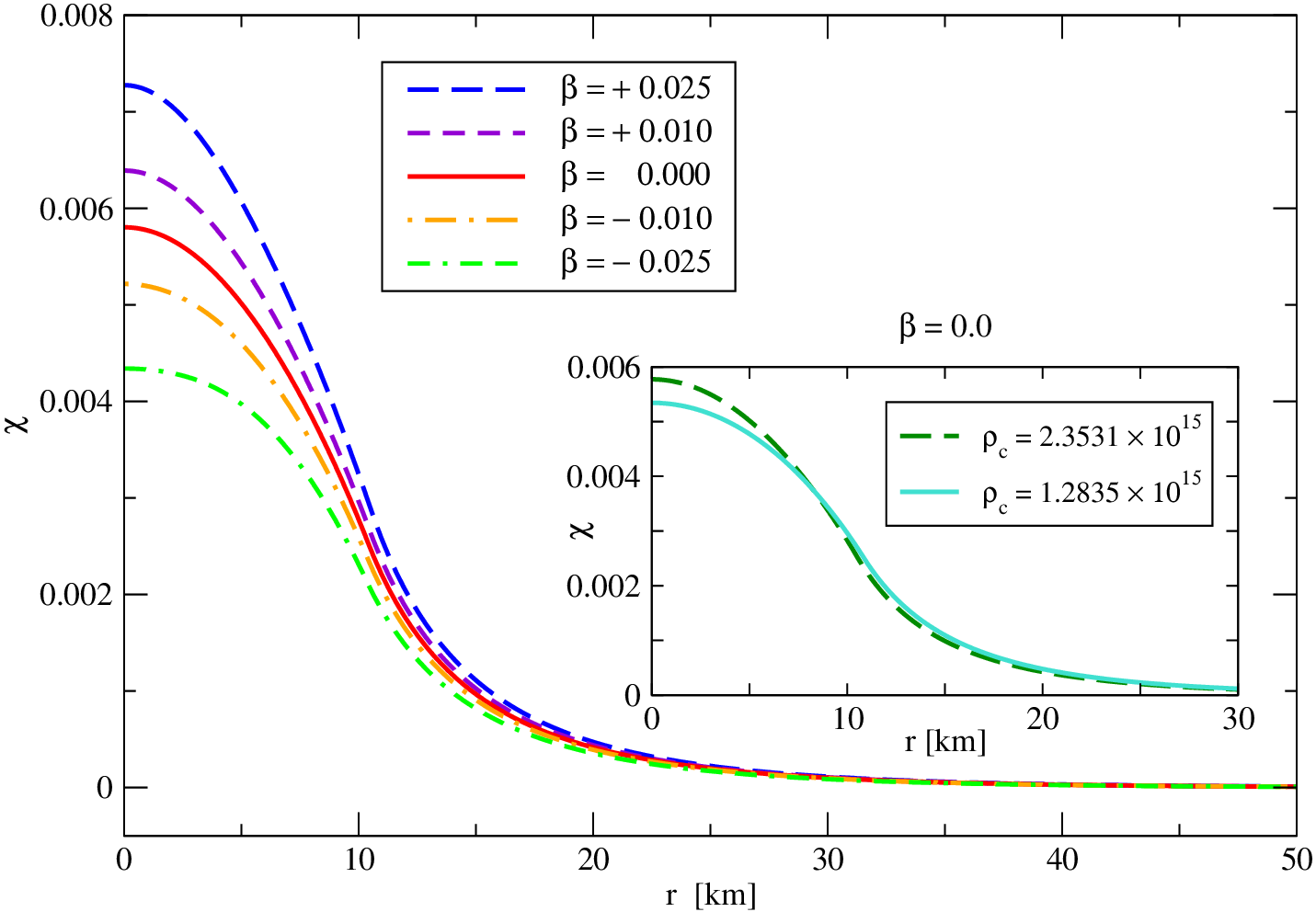}
\caption{Radial profile of scalar curvature $R(r)$ for different values of $\beta$ ($\beta=4B\omega$) with central density $\rho_c = 2.5414\times 10^{15}$ g cm$^{-3}$.  The inset shows the scalar curvature profile for the Starobinsky model ($\beta=0$) with two different central densities $\rho_c$ (in g cm$^{-3}$).}
\label{Figure_2}
\end{figure}
Figure \ref{Figure_1} plots pressure $p$ as a function of the radial coordinate $r$ for the central density $\rho_c = 2.5414\times 10^{15}$ g /cm$^{3}$ for $\beta =0.025$,
$0.01$, $-0.01$ and $-0.025$. We see that the magnitude of pressure gradient increases (decreases) with respect to the pure Starobinsky case ($\beta=0$) for positive (negative) value of $\beta$ (or $\omega$) due to the additional term in the extended TOV equation (\ref{MSG_Eq4}). This in fact pushes (pulls) the stellar boundary outward (inward) as compared with the  pure Starobinsky case.  The slight increase (decrease) in stellar radius $r_s$ for positive (negative) values of $\beta$ (or $\omega$) can be seen in the inset of Figure \ref{Figure_1}.

Radial profiles of the scalar curvature $R(r)$ for different values of $\beta$ with central density $\rho_c = 2.5414\times 10^{15}$ g /cm$^{3}$ are shown in Figure \ref{Figure_2}. It may be observed that the choice of the central scalar curvature $R_c$ is strongly correlated with $\beta$ (or equivalently $\omega$) in that the value of $R_c$ increases with increasing values of $\omega$. On the other hand,  when we fix $\beta=0$ and vary $\rho_c$, the central value $R_c$ is found to be higher for a higher value of $\rho_c$ as shown in the inset of Figure \ref{Figure_2}. Although there is an increase in the value of $R_c$ when $\rho_c$ is increased, the scalar curvature $R$ falls off rapidly for higher value of $\rho_c$ than for the lower one. In the former case (with fixed $\rho_c$  and varying $\beta$),  the Ricci scalar maintains higher values thoughtout the star for higher value of $\beta$.This is consistent with the fact that the perturbative $\omega$ terms in Eq.(\ref{MSG_Eq1}) add with the term $\frac{\kappa}{6\alpha} T e^{\lambda}$, so that the effective value of $T$ changes which is equivalent to a changed value of matter content with respect to the Starobinsky case. Since these perturbative terms disappear outside the star, they act as if they were an additional matter content in Starobinsky gravity.

\begin{figure}[h!]
\centering
\includegraphics[width=0.48\textwidth]{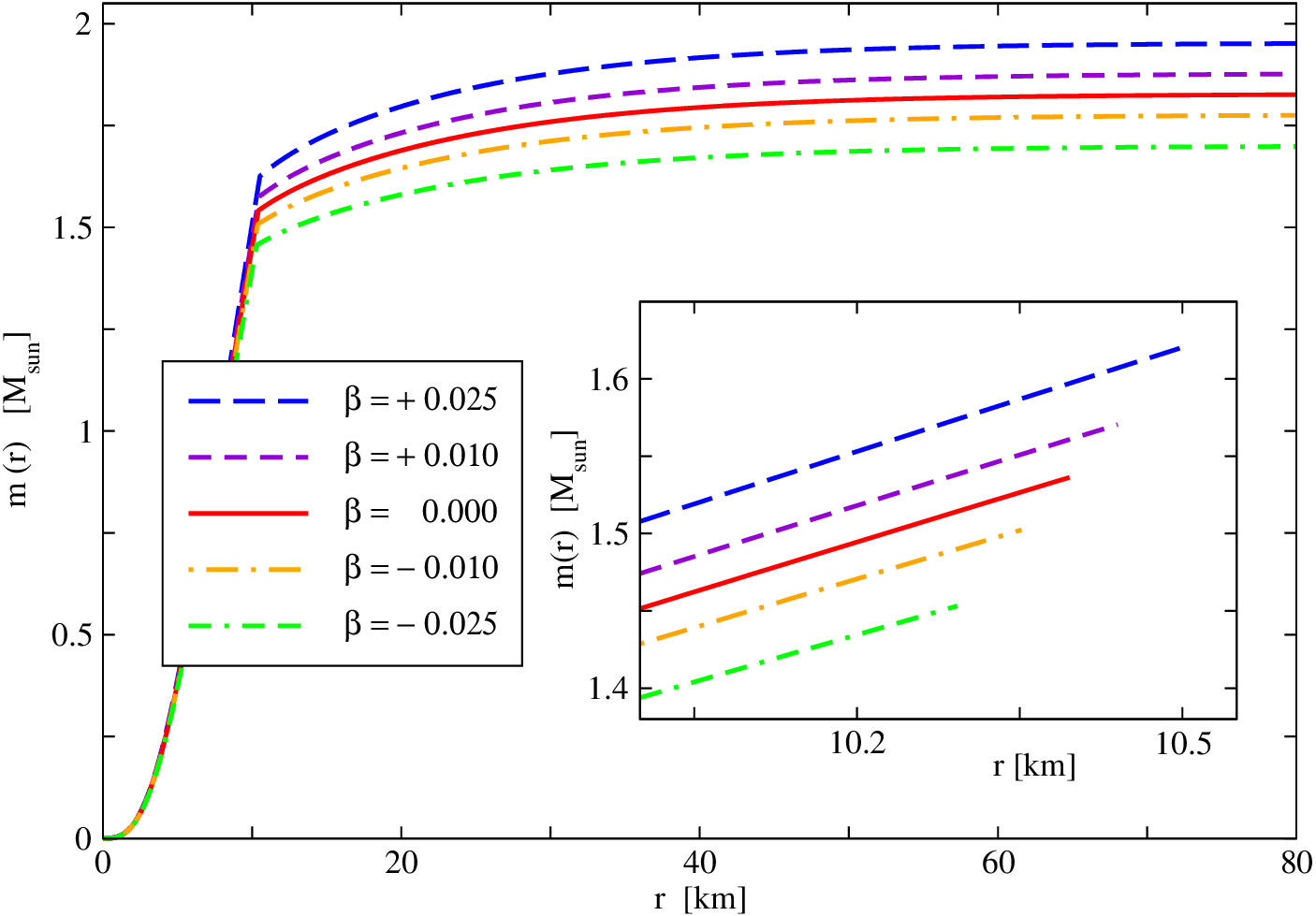}
\caption{Radial profile of total mass $m(r)$ for different values of $\beta$ ($\beta=4B\omega$) with central density $\rho_c = 2.5414\times 10^{15}$ g cm$^{-3}$. The inset shows the interior mass profile up to the stellar radius $r_s$. }
\label{Figure_3}
\end{figure}

Figure \ref{Figure_3} shows the total mass profile $m(r)$ with $\rho_c = 2.5414\times 10^{15}$ g /cm$^{3}$ for different values of $\beta$.  The inset represents the mass profile up to the stellar surface $r=r_s$. In comparison with Starobinsky gravity ($\beta=0$), we observe a slight increase (decrease) in the stellar mass $M_s$ and stellar radius $r_s$ when $\beta$ is positive (negative). Further we observe that these changes in stellar mass $M_s$ and radius $r_s$ are larger for $\beta=\pm0.025$ than $\beta=\pm0.01$. On the other hand, we  see that the  mass $M$ measured by a distant observer increases appreciably with increasing $\beta$. 
 
As seen from Figure \ref{Figure_2},  the scalar curvature does not decrease to zero at  the surface of the star and it falls off outside the star. This fall-off is similar to a Yukawa function (as shown in Section \ref{Far_field_sol}). There is a gravitational mass contribution due to the non-vanishing scalar curvature outside the star. The mass profiles shown in Figure \ref{Figure_3} contains both contributions, stellar plus gravitational. The inset shows only the stellar mass contribution that does not extend beyond the stellar radius $r_s\sim10$ km. The main graphs in Figure \ref{Figure_3} show both contributions (stellar plus gravitational) extending beyond $r_s\sim10$ km. We see that the combined mass profiles approach asymptotic values for large $r$ ($\sim60$ km).  A sufficiently distant body experiences the gravitational field of the combined mass.

\subsection{Mass-radius relations}

In this section we study the mass-radius ($\mathcal{M}-\mathcal{R}$) relations obtained from the field equations for a continuous range of central density $\rho_c$ or equivalently the central Ricci scalar $R_c$. In addition, we verified that all energy conditions are satisfied. 

\begin{figure}[h!]
\centering
\includegraphics[width=0.48\textwidth]{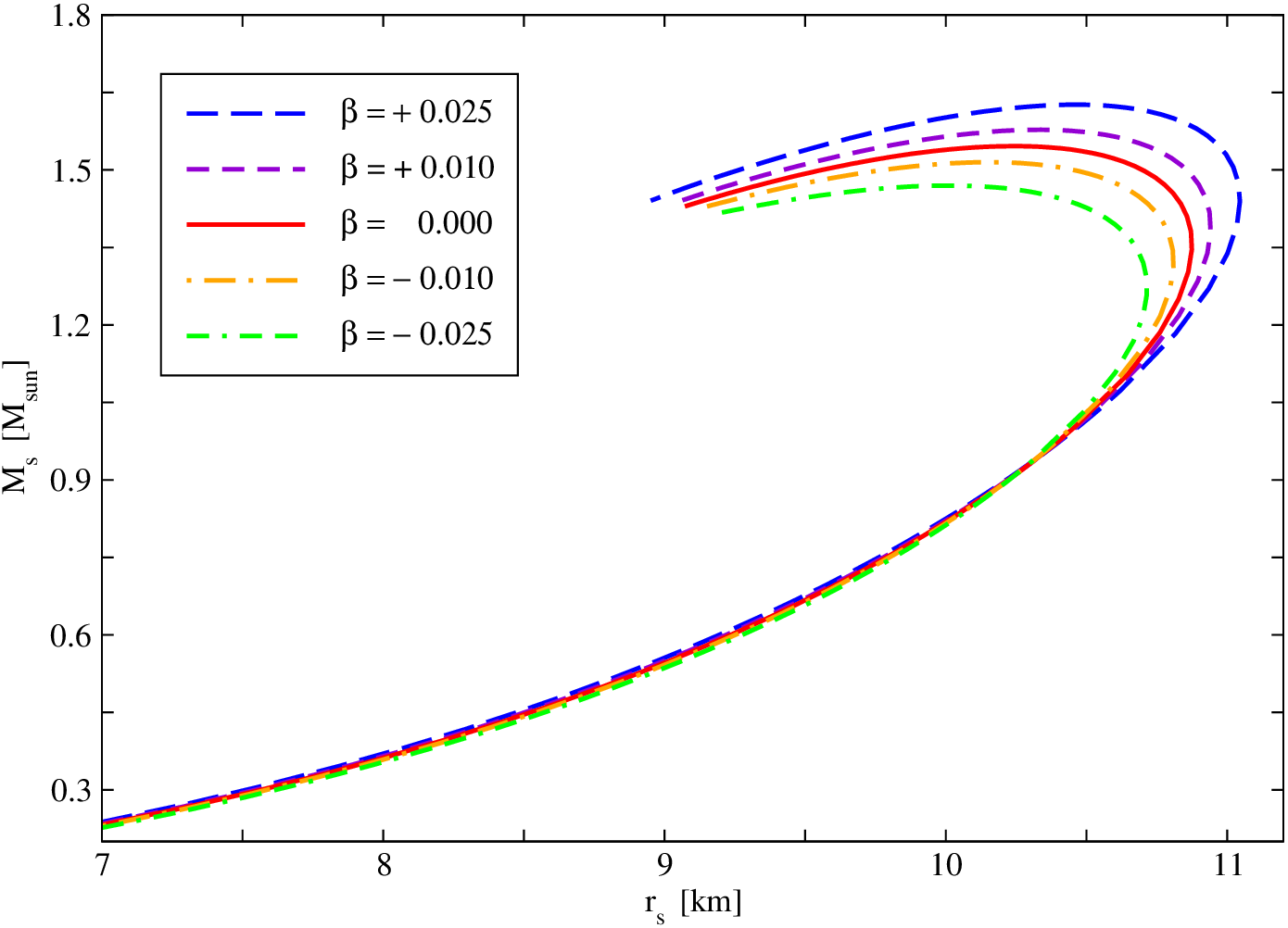}
\caption{Mass-radius relation (between stellar mass $M_s$ and stellar radius $r_s$) for different values of $\beta$. }
\label{Figure_4}
\end{figure}

Fig. \ref{Figure_4} represents the relation between stellar mass $M_s$ and stellar radius $r_s$ for different values of $\beta$. We see that for a particular value of $M_s$, $r_s$ increases with increase in $\beta$ in the higher mass regime.  In the same regime, if we fix $r_s$, $M_s$ is found to increase with increasing $\beta$. This fact signify that the presence of the $\omega$ terms strengthens the effect of gravity to balance the increased pressure gradient as inferred from the pressure profiles studied in Section \ref{I_E_sol}.

\begin{figure}[h!]
\centering
\includegraphics[width=0.48\textwidth]{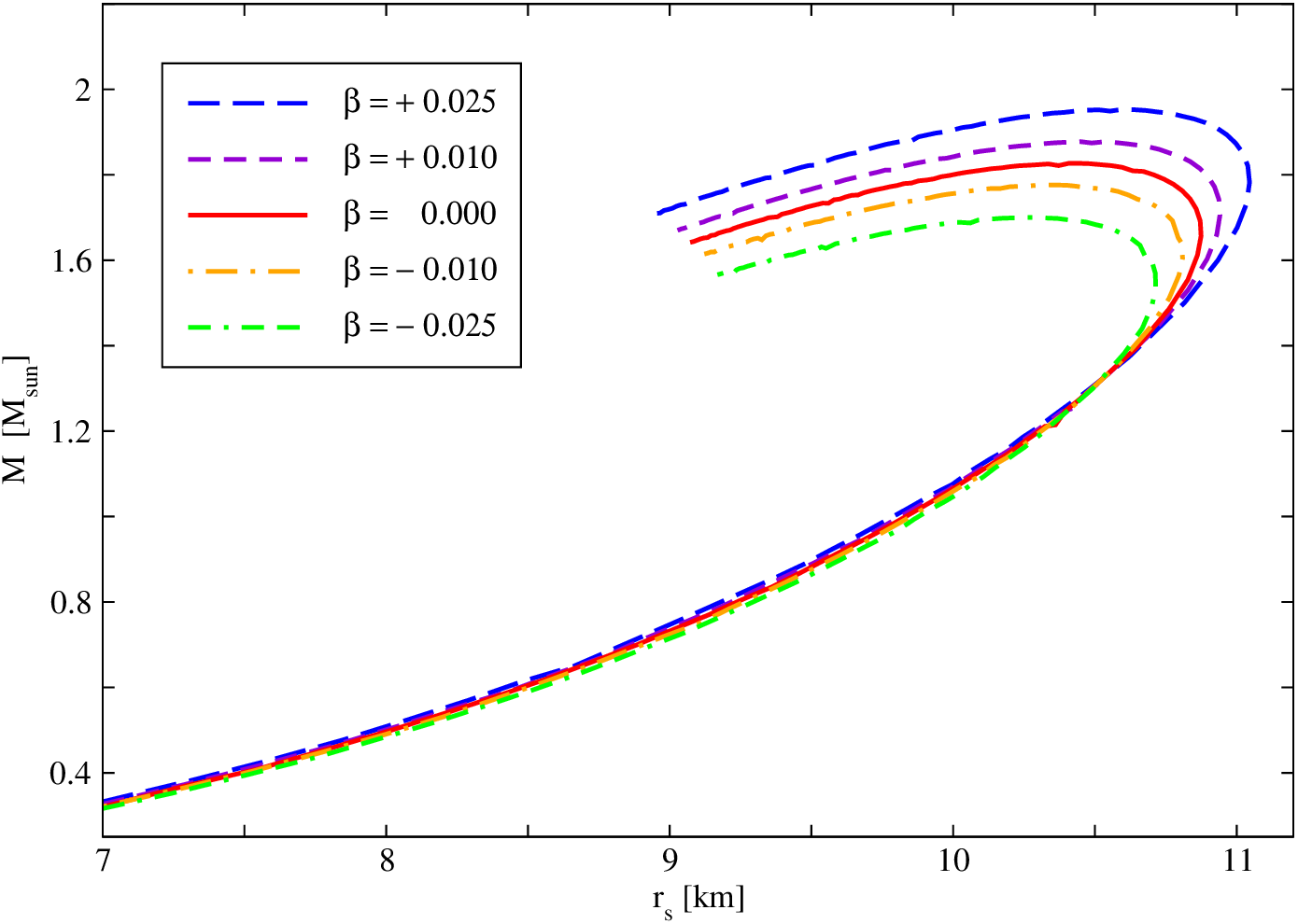}
\caption{Total mass $M$ measured by a distant observer versus stellar radius $r_s$ for different values of $\beta$.}
\label{Figure_5}
\end{figure}

\begin{figure}[h!]
\centering
\includegraphics[width=0.48\textwidth]{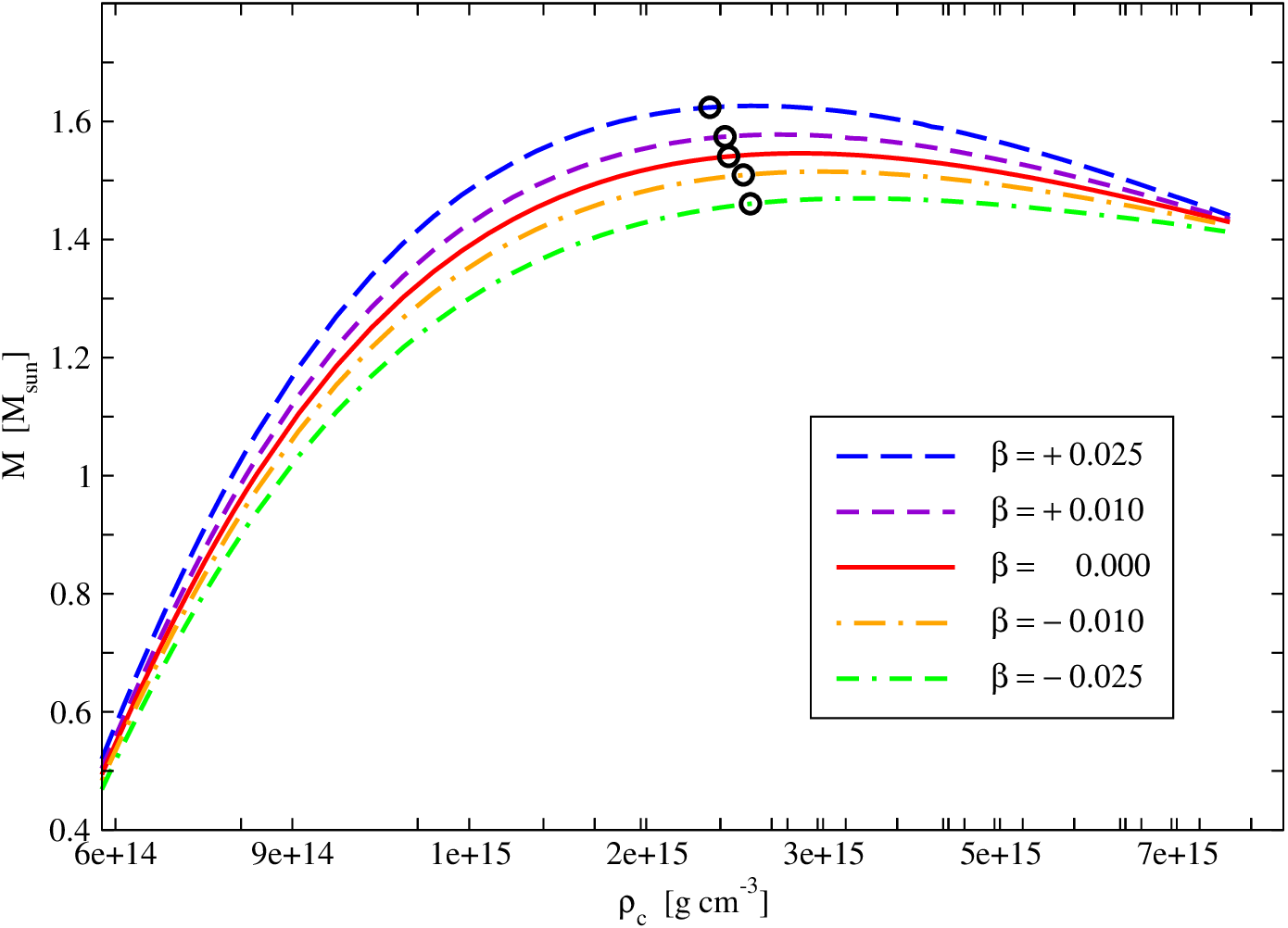}
\caption{Stellar mass $M_s$ versus central density $\rho_c$ for different value of $\beta$, where the mass $M_s^\ast$ is shown by open circle.}
\label{Figure_6}
\end{figure}

Figure \ref{Figure_5} presents the mass $M$ measured by a distant observer against the stellar radius $r_s$ for different values of $\beta$. As noted earlier, the mass $M$ consists of contributions from both the stellar mass and non-vanishing scalar curvature extending beyond the stellar radius $r_s$. This mass was calculated up to a sufficiently high radial distance ($r_{\rm max}$) until the scalar curvature approached very close to zero with the condition given by Eq.~(\ref{M}).  The relationship between the curves in Figure \ref{Figure_5} bear similarity with those in Figure \ref{Figure_4} giving qualitatively similar conclusion. However we note the important fact that maximum observed mass $M^\ast$ are appreciably higher than those in Fig. \ref{Figure_4}.

\subsection{Stability and energy conditions}

In this section, we study mass versus central density to find the maximal mass from the stability of equilibrium configurations. The stable configuration corresponds to the region in the mass-central density curve where $\frac{\partial M}{\partial \rho_c}>0$, whereas the unstable region is given by $\frac{\partial M}{\partial \rho_c}<0$  \cite{Shapiro_BHbook}. The onset of instability is identified as the point where $\partial M/\partial \rho_c=0$, and the mass corresponding to this point is the maximal. To study the stability, we first examine stellar mass $M_s$ versus central density $\rho_c$ in Fig. \ref{Figure_6} for different values of $\beta$. We see that the maximal stable mass $M_s^\ast$ (corresponding to the maximal total mass $M^\ast$) increases as $\beta$ increases from $\beta=-0.025$ to $\beta=+0.025$.

\begin{figure}[h!]
\centering
\includegraphics[width=0.48\textwidth]{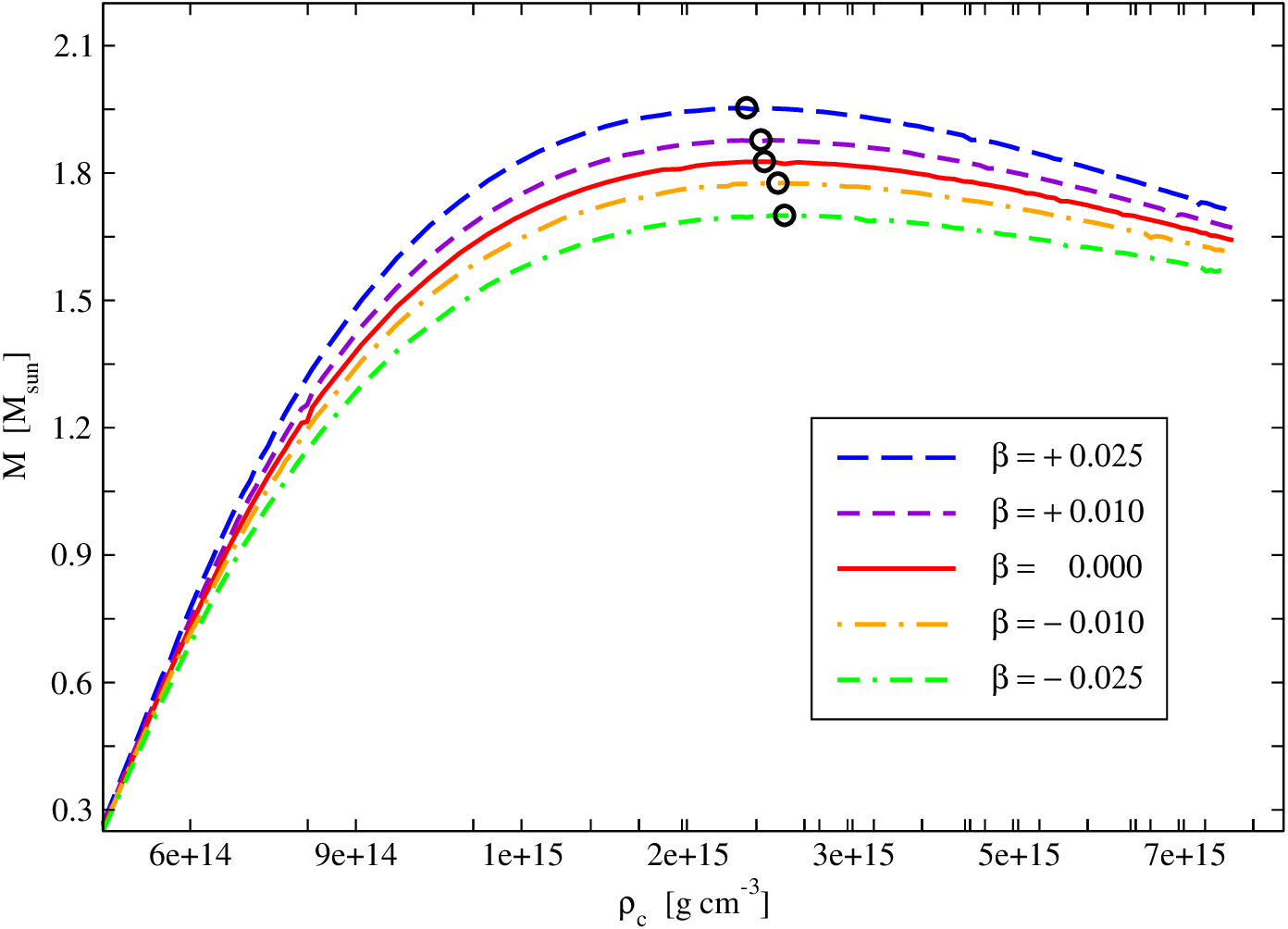}
\caption{Total mass $M$ versus central density $\rho_c$ for different value of $\beta$, where the maximal mass $M^\ast$ is shown by open circle}
\label{Figure_7}
\end{figure}

A similar trend is observed in Fig. \ref{Figure_7}, where the mass $M$ observed by a distant observer is plotted against the central density $\rho_c$. Here we see that the maximal mass $M^\ast$ (denoted by the open circle in the figure) shift to appreciably higher value with respect to the stellar values $M_s^\ast$. 

Table \ref{Table_1} displays maximal mass values $M^\ast$ observed by a distant observer for different values of $\beta$. The corresponding stellar mass values $M_s^\ast$, stellar radius $r_s^\ast$, central Einstein Ricci scalar $^ER_{c}^\ast$, central Starobynsky Ricci scalar $R_{0c}^\ast$, central Ricci scalar $R_{c}^\ast$,  and central density $\rho_c^\ast$ are also displayed. We see that the maximal mass value $M^\ast$ increases and approaches $\sim$ 2 M$_\odot$ as $\beta$ is increased. We note that this increase is appreciable even for very small magnitudes of $\beta$, suggesting a measurable effect played by gravity-matter interaction.

We verify the validity of the perturbative results by estimating the maximal value of $\omega T$ corresponding to the maximal mass. For $\beta=0.025$ (or $\omega=0.025/4B$) and central density $\rho_c^\ast= 2.4173\times10^{15} $ g/cm$^3$, we get $|\omega T_c| = (1-3k)\beta\rho_c c^2/4B + 3\beta k  = 4.36\times10^{-2}$. Thus the maximum value of $\omega T$ is of the order of $10^{-2}$, giving assurance to the validity of the perturbative results.

\begin{figure}[h!]
\centering
\includegraphics[width=0.48\textwidth]{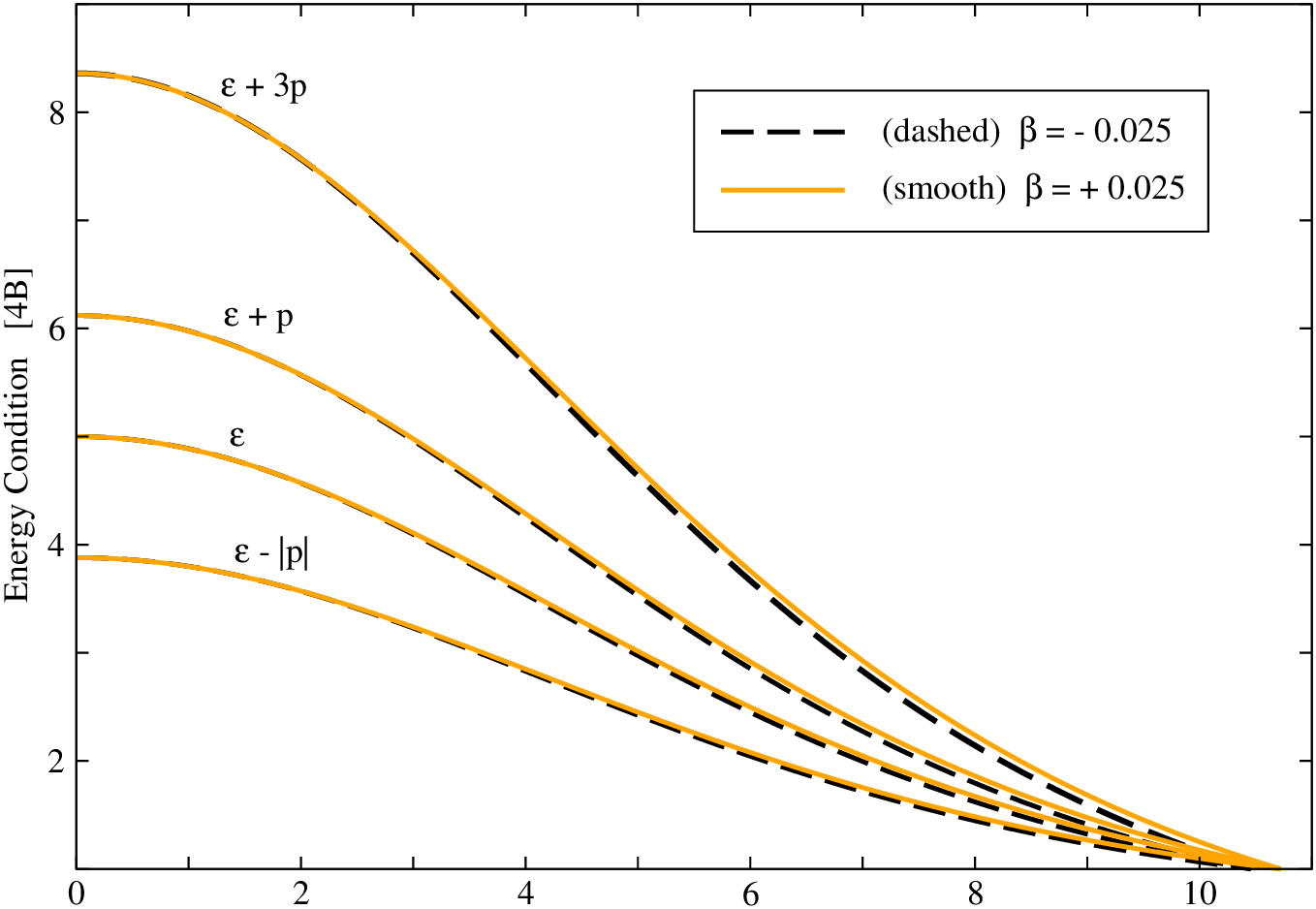}
\caption{Radial profiles of  energy conditions (ECs): null energy condition ($\varepsilon\geq0$), weak energy condition ($\varepsilon+p\geq0$), strong energy condition ($\varepsilon+3p\geq0$), and dominant energy condition ($\varepsilon-|p|\geq0$).}
\label{Figure_8}
\end{figure}

Figure 8 displays the energy conditions \cite{Francisco2019}, namely, null energy condition ($\varepsilon\geq0$), weak energy condition ($\varepsilon+p\geq0$), strong energy condition ($\varepsilon+3p\geq0$), and dominant energy condition ($\varepsilon-|p|\geq0$). We see that all energy conditions are satisfied because they are positive in the entire region of the star. These energy conditions are valid to a good approximation since they are large throughout the star (lying between  $\sim1$ and $\sim10$ in the units of $4B$) compared to the highest perturbation ($\sim 10^{-2}$ at the centre).

\section{Discussion}\label{Discussion}

\begin{table*}
\centering
\caption{Maximum stable mass $M^\ast$ for different values of $\omega$. The corresponding values of stellar mass $M_s^\ast$, stellar radius $r_s^\ast$, central Einstein Ricci scalar $^ER_{c}^\ast$, central Starobynsky Ricci scalar $R_{0c}^\ast$, central Ricci scalar $R_c^\ast$ and central density $\rho_c^\ast$ are also displayed.}
\begin{tabular}{cccccccc} 
\hline\\[-3ex] \hline\\[-2.0ex]
$\beta$ & $\rho_c^\ast$ (g cm$^{-3}$) & $^ER_c^\ast$ ($\times10^{-2}r_g^{-2}$) &$R_{0c}^\ast$ ($\times10^{-3}r_g^{-2}$) &$R_c^\ast$ ($\times10^{-3}r_g^{-2}$) & $r_s^\ast$ (km) & $M_s^\ast$ (M$_\odot$) & $M^\ast$ (M$_\odot$) \\[0.5ex]
\hline \\[-2.5ex]
$-\ 0.025$    & $2.6526\times10^{15}$  &  3.189908	& 5.81876  &	4.286802	&   10.2686	&   1.46058	&	1.70031 \\
$-\ 0.01$ 	   & $2.6098\times10^{15}$   &	3.162040  & 5.81296  &	5.209743	&   10.3482	&   1.50937     &	1.77659 \\
$0.0$     	   & $2.5242\times10^{15}$   &  3.106321	& 5.80072  &	5.800721	&   10.4265	&   1.54080	&	1.82725 \\
$0.01$   	   & $2.5029\times10^{15}$   &	3.092391	& 5.79744  &   	6.375852	&   10.4790	&   1.57427	& 	1.87787 \\
$0.025$ 	   & $2.4173\times10^{15}$   &  3.036672	& 5.78384	 &     7.180354  &   10.5851	&    1.62390	&	1.95371 \\[0.5ex]
\hline\\[-3ex] \hline
\end{tabular}
\label{Table_1}
\end{table*}

\begin{figure}[h!]
\centering
\includegraphics[width=0.48\textwidth]{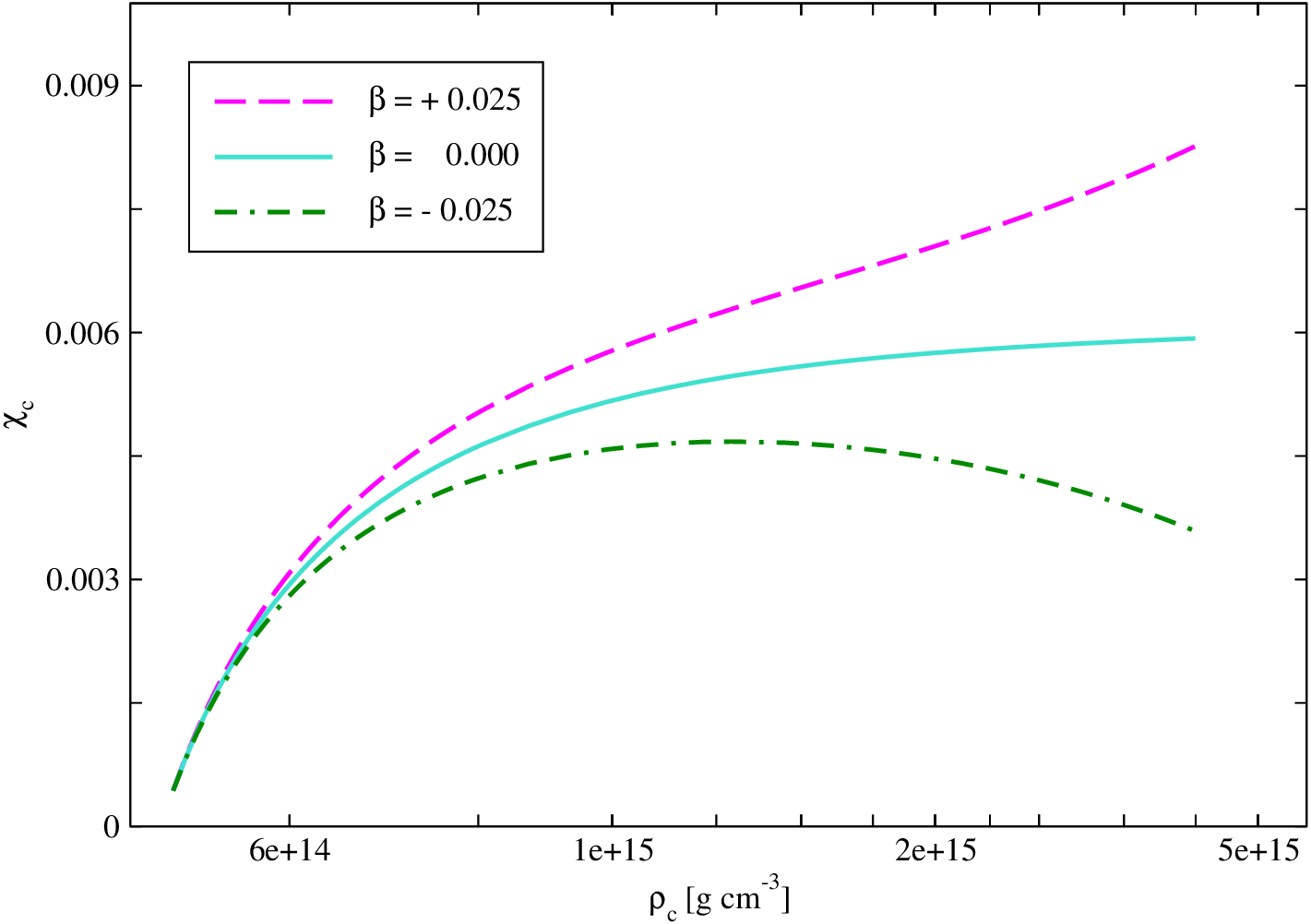}
\caption{Central value of Ricci scalar $\chi_c$ ($=R r_g^{2}$) versus central density $\rho_c$ for different values of $\beta$, namely, $\beta=\pm0.025$ and $\beta=0$.}
\label{Figure_9}
\end{figure}

In the original case of Einstein's gravity,  the Ricci scalar is a linear function of the central density, expressed as $\chi=\kappa r_g^2 \{(1-3k)\rho c^2+12kB\}$. One might expect such a linear relationship in the Starobinsky model in the low density regime where $\alpha R\ll1$. Besides, the same behaviour is expected in the present model  for low central densities where the term $R$ dominates over $\alpha R^2$ and $\omega R T$. However, the situation is completely different in the high density regime in both Starobinsky and the present model. In the Starobinsky model, we found that at higher central densities, the central Ricci scalar $R_{0c}$ varies slowly as a function of $\rho_c$, as shown in Figure~\ref{Figure_9}. On the other hand, in the present model, depending on the sign of $\beta$ (or $\omega$), we find that the central Ricci scalar $R_c$ would either increase or decrease with respect to the Starobinsky case. Figure~\ref{Figure_9} shows the variation of central Ricci scalar with respect to central density for different values of $\beta$. For positive $\beta$, we see that the $R_c$ value increases compared to the Starobinsky model in the higher density regime. This was expected since the additional terms in the present model contribute at higher densities as previously noted in Section~\ref{I_E_sol}. The opposite is true for the case of negative $\beta$ values where we found that the central Ricci scalar $R_c$ values lie below the Starobinsky values for higher densities as shown in Figure~\ref{Figure_9}. This happens because $\mathcal{O}(\omega)$ term gives a negative contribution in this case. 

For any positive $\beta$, the curve in Figure~\ref{Figure_9} lies above the Starobinsky case, so that maximum mass values higher than the Starobinsky case would be obtained. On the other hand, the curve for any negative $\beta$ lies below the Starobinsky case implying that the maximum mass values lower than the Starobinsky case would be obtained.

It may be recalled from the discussion in Introduction that, in the Starobinsky gravity ($\omega=0$), the star is surrounded by a gravitational halo since the Ricci scalar is non-vanishing outside the star. In fact we see the same behaviour from Figure \ref{Figure_2} in the present model ($\omega\neq0$) as well. The far-field solution (given by equation \ref{Vaccum_sol}) shows that the Ricci scalar (and the gravitational halo) falls off exponentially as $r\rightarrow\infty$.
 \begin{figure}[h!]
\centering
\includegraphics[width=0.48\textwidth]{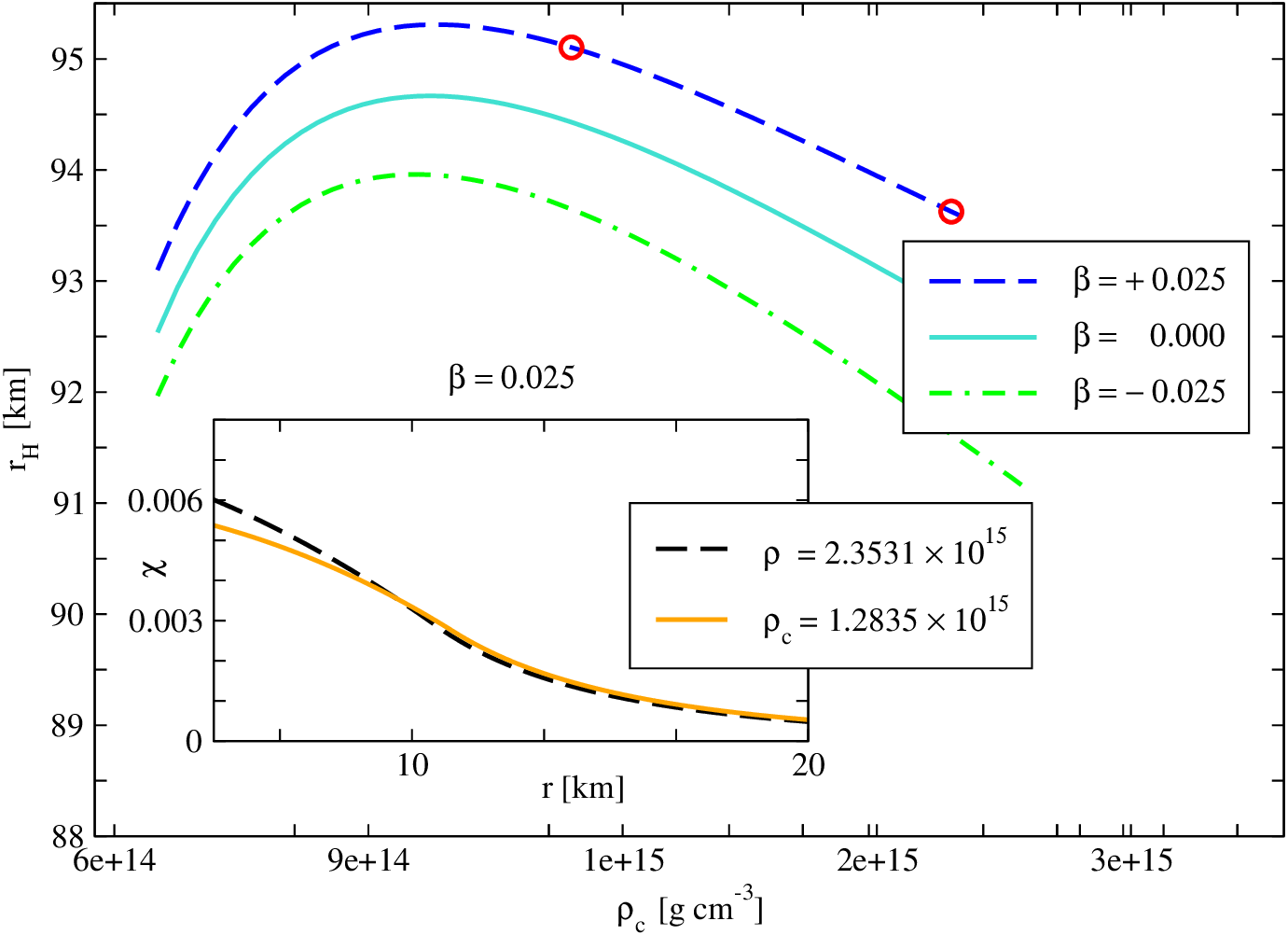}
\caption{The radius of gravitational halo  $r_{\rm H}$ versus the central density $\rho_c$ until the onset of gravitational instability at $\rho_c^\ast$. The inset shows that the radial profiles for the scalar curvature $\chi$ (corresponding to the points denoted by the open circle in the main graph) falls off faster for a higher density, similar to the Starobinsky case ($\beta=0$), as shown in the inset of Figure \ref{Figure_2}. }
\label{Figure_10}
\end{figure}
Moreover, it is evident from Figure \ref{Figure_2} that this fall-off is faster for higher values of the central density $\rho_c$. Figure \ref{Figure_10} plots an effective radius $r_{\rm H}$ of the gravitational halo (defined by $\chi(\eta_{\rm H})=10^{-7}$) with respect to the central density $\rho_c$ until the onset of gravitational instability at $\rho_c^\ast$. (Densities beyond the threshold $\rho_c^\ast$ are outside the scope of the present theory for equilibrium configurations.) It is seen from the right-hand part of the graphs in Figure \ref{Figure_10} that the radius of the halo $r_{\rm H}$ decreases with increasing central density $\rho_c$.  At the same time, the value of  $r_{\rm Sch}=\frac{2GM}{c^2}$ increases with increasing central density $\rho_c$, as seen from Figure \ref{Figure_7}. These two opposite behaviours (shrinking and expansion) continue as $\rho_c$ increases. Thus it is apparent that, when the star collapses to a black hole (with an infinite central density), the gravitational halo would shrink and will be well inside the horizon, leading to a vanishing Ricci scalar outside the horizon. This scenario is consistent with the fact that when the coefficient of $R^2$ term is positive, the only static spherically symmetric solution of a black hole with a regular horizon is the Schwarzschild solution, as shown in Ref. \cite{Mignemi1992}.

\section{Conclusion}\label{Conclusion}

In this paper, we considered a form of $f(R,T)$ gravity that includes a coupling between gravity and matter on the background of the Starobinsky model. While the Straobinsky model takes account of quantum fluctuations \cite{Starobinsky1980} and $f(R,T)$ gravity may arise due to quantum effects \cite{Harko2011}, a coupling between matter and gravity is expected to bring about the features in quark stars where the gravitational field is extremely strong. In particular,  the stellar structure of quark stars, with equation of state coming from the bag model, is expected to undergo an measurable change due to this coupling. Moreover, we speculate that the maximum mass limit would change appreciably so that astrophysical observations on binary pulsars could be given a theoretical basis. 

The Starobinsky model has been applied to quark stars by other authors \cite{Astashenok2015,Astashenok2017} to find their stellar structure. This produced a different stellar structure from the pure GR case and it was found that the maximum mass limit increased from the GR case due to an additional contribution from gravitational mass enveloping the stellar mass. In the present case, we find that this mass is further increased due to additional contribution from the coupling between gravity and matter (for positive values of $\omega$). 

To assert the above features, it was sufficient to treat the gravity-matter coupling as a perturbation keeping in mind that the coupling constant is sufficiently small for the validity of the perturbation treatment. We adopted this perturbation treatment in the background of unperturbed solutions of the Starobinsky case. Remarkably, such a treatment gives physically acceptable solutions for both signs of the coupling constant $\omega$ representing the strength of gravity-matter interaction.

The gravity-matter coupling term increases (decreases) the magnitude of the pressure gradient $p'(r)$ for positive (negative) values of $\omega$ pushing (pulling) the stellar boundary outward (inward) as compared to the pure Starobinsky case. Moreover the strength $\omega$ of the gravity-matter coupling determines the central value of the scalar curvature $R_c$ for a given central density $\rho_c$.  The scalar curvature maintains higher (lower) values throughout the star compared to the pure Starobinsky case for positive (negative) values of $\omega$ as the effective matter content increases (decreases) within the star. It is interesting to see that, although there is small increase (decease) in the stellar mass $M_s$ for positive (negative) values of $\omega$, the gravitational mass contribution enveloping the star increases (deceases) appreciably with respect to the Starobinsky case. This is because of the increased (decreased) scalar curvature exterior to the star contributing a greater (lesser) gravitational mass than the Starobinsky case. Consequently the quark star can support higher values of maximal total mass ($M^\ast$) than the Starobinsky case for positive values of $\omega$. 

Recent observations of binary millisecond pulsars, have yielded the pulsar masses to be $\sim2$ M$_\odot$ \cite{Antoniadis2013,Demorest2010,Fonseca2016,Cromartie2019}. Such a high value of mass cannot be explained by models based on hyperon or boson condensate equations of state for neutron stars, leading to their possibility of being quark stars. We see that our present model with gravity-matter coupling, although treated as a perturbation, is capable of supporting high values of masses of quark stars.

{\bf{Acknowledgements}}
Arun Mathew is indebted to the Indian Institute of Technology Guwahati for financial assistance through a Research and Development fund. The authors would like to thank the reviewer for constructive comments and suggestions that improved the presentation in the paper.


\end{document}